# Spin relaxation in low dimensional systems

L. Viña

Departamento de Física de Materiales C-IV, Universidad Autónoma de Madrid,
Cantoblanco, E-28049 Madrid, Spain

**Abstract**

We review some of the newest findings on the spin dynamics of carriers and excitons in GaAs/GaAlAs quantum wells. In intrinsic wells, where the optical properties are dominated by excitonic effects, we show that exciton-exciton interaction produces a breaking of the spin degeneracy in two-dimensional semiconductors. In doped wells, the two spin components of an optically created two-dimensional electron gas are well described by Fermi-Dirac distributions with a common temperature but different chemical potentials. The rate of the spin depolarization of the electron gas is found to be independent of the mean electron kinetic energy but accelerated by thermal spreading of the carriers.

Author to be contacted: Luis Viña, Departamento de Física de Materiales, C-IV-510.
Universidad Autónoma de Madrid. E-28049 Cantoblanco, Madrid. SPAIN.
e-mail: luis.vina@uam.es

## 1. Introduction

Useful electronic devices rely on the precise control of electronic charge, and in general the fact that the electrons also have a spin is ignored. However, the scattering processes for electrons depend on their spin state. Spin related effects of carriers are attracting a lot of attention in different fields of condensed matter physics. Recently, interest in electronic spin polarization in solid-state systems has grown fuelled by the possibility of producing efficient photoemitters with a high degree of polarization of the electron beam, creating spin memory devices and spin transistors as well as exploiting the properties of spin coherence for quantum computation. The determination of the spin-flip rates is extremely important for electronic applications, because if the spins relax too rapidly, the distances traveled by spin-polarized currents will be too short for practical applications.

Studies on spin-polarized transport are growing dramatically especially in ferromagnetic metals, which present a splitting between the up and down spin states ("exchange splitting") to lower the total energy of the system by avoiding the high energy of having a high density of states at the Fermi level. Therefore the current-carrying electrons at the Fermi level should be polarized. However, in tunneling experiments from a ferromagnetic metal film through a nonmagnetic insulating barrier into a superconducting metal film, it was found surprisingly that the percentage polarization of the current did scale with the total moment, given by the net polarization of the electrons, and not with the polarization of the electrons at the Fermi level.[1]

Devices such as a magnetic valve, which exploits the dependence of spin-polarized transport upon the spin-dependent density of states available at the Fermi level in two ferromagnetic metal films, have been demonstrated. These devices are analogous to optical polarizers; however the minimum transport is obtained when the magnetic moments of the two magnetic films are rotated 180º away from the parallel, whereas for the optical case the minimum transmission is obtained for 90º orientation of the polarizer axis.[1] Very recently, Monsma et al. have established that spin-dependent scattering of electrons can be exploited in a metal-base transistor-like structure.[2] This can be obtained by launching hot electrons from a semiconductor emitter into a very thin metallic magnetic base. An all-optical gate switch, which uses the polarization-dependent optical nonlinearity of 2D excitons in QW's, has been proposed and high-speed picosecond switching has been demonstrated,[3] and photon-spin controlled lasing oscillations in GaAs surface-emitting lasers at room temperature have been reported.[4]

There are various experiments one can do to detect optical orientation, such as spin polarization of electron photoemission, electron paramagnetic resonance, nuclear magnetic resonance, spin-dependent transport, Faraday rotation, and spin-dependent pump-and-probe or recombination. The Hanle effect,[5] i.e., the spin depolarization of carriers subject to a transverse magnetic field, has been traditionally used to obtain quantitative values for the spin relaxation times.[6-8] Optical pumping, with linearly polarized light, and orientation, with circularly polarized light, of carrier spins is a powerful method to investigate relaxation processes in semiconductors and has also found applications in spin-polarized electron sources. With the appearance of ultrafast lasers, the spin relaxation time can be also measured in a direct way. Ultrafast time-resolved photoluminescence (TR-PL) has proven to be a very useful tool to probe carrier dynamics in semiconductors. The emitted luminescence, after an excitation with a light pulse, reflects the temporal evolution of the carrier distribution and can be analyzed to study energy as well as orientation relaxation rates.[9]

In bulk semiconductors, the spin relaxation of excitons and free carriers has been extensively studied in the past.[9-12] Most of the investigations have dealt with the study of conduction-band electrons in p-type materials,[13-18] or excitons in intrinsic ones.[19-21] A polarization analysis of the band-to-band luminescence was performed to obtain the various energy, momentum and spin-relaxation mechanisms for electrons.[22, 23] The amount of work has been more limited on n-type materials.[7, 8, 24-26] The experimental work was stimulated by theoretical studies of spin relaxation processes by Elliot and Yafet,[27, 28] Dyakanov' and Perel'[29-31] and Bir et al.,[32] and different mechanisms responsible for the spin flip processes have been identified.[9, 15, 17, 33, 34] The dependence of the various spin-relaxation mechanisms on temperature and doping has been used to distinguish between them.[35-37]

The relative importance of these mechanisms is modified in semiconductor quantum wells (QW's) due mainly to three reasons: a) the new electronic bandstructure, which affects the hole spin-relaxation; b) the enhancement of the excitonic effects, leading to a larger spin-flip rate due to exciton exchange; and c) the higher carrier mobility, which is important for mechanisms sensitive to carrier momentum relaxation, such as the D'yakanov-Perel' (DP) one.[29] Experimental investigations on the spin dynamics in low-dimensional semiconductors have flourished in the last decade. Many works deal with the spin processes of excitons,[38-60] including indirect excitons in type II QW's,[61] excitons in zero-dimensional quantum disks,[62] the influence of external electric fields,[63] and the inhibition of spin relaxation by a fast energy relaxation process (emission of a longitudinal-optical phonon).[64] Fewer investigations deal with the spin-flip of individuals electrons and holes in 2D systems.[45, 48, 65-71] Extensive theoretical studies have been also done on the spin-flip relaxation of free carriers, electrons and holes,[72-79] and excitons.[80-87]

In quasi-two-dimensional systems the exciton energy is renormalized to higher values at high densities.[88, 89] A model of Schmitt-Rink et al. predicted this effect as a consequence of a strong reduction of the long-ranged Coulomb correlation interaction in a 2D system.[90] This is in strong contrast with the situation for three-dimensional excitons, which maintain their energies even at high densities. This effect is due to a compensation between two many-body effects: a repulsive contribution originating in the Pauli exclusion principle acting on the fermions forming the excitons, and an attractive one, similar to a van der Waals interaction. The latter is precisely reduced in 2D systems giving rise to the blue shift of the excitons with increasing density.

An additional effect has been observed when excitons with a well defined spin are created.[45] Using circularly polarized ($\sigma^+$) light to excite the samples, close to resonant formation of heavy-hole excitons, an energy splitting between the two components, $\sigma^+$ and $\sigma^-$, of the heavy-hole (hh) exciton luminescence has been reported. The $\sigma^+$ component is always at higher energies than that of $\sigma^-$ helicity. This effect has also been confirmed by other TR-PL studies[51, 70, 91-93] and by femtosecond pump and probe measurements in GaAs QW's under an external magnetic field.[94] A model has been recently developed to explain it as arising from inter-excitonic exchange interaction.[95, 96]

Some other manifestations of spin-dependent exciton-exciton interactions have been also shown recently.[57, 70] They include a spin-dependent optical dephasing time and a linewidth difference between the two photoluminescence components $\sigma^+$ and $\sigma^-$ (the luminescence component co-polarized with the laser excitation is narrower than the counter-polarized one). All these can be interpreted as a result of interexcitonic exchange.

Exciton, as well as individual electron and hole, spin dynamics have been also investigated, in the presence of an external magnetic field, on heterostructures based on diluted semimagnetic semiconductors such as $Zn_{1-x}Mn_xSe$, using ultrafast Faraday spectroscopy[97-100] and TR-PL,[101] or $Cd_{1-x}Mn_xTe$.[102] A very rich variety of spin phenomena, absent in traditional semiconductor heterostructures, have been found in the Faraday rotation experiments, and the important role played by correlation effects between excitons of different spin has been recognized.[103] Some of the results of these experiments seem to be sample dependent: while in the $Cd_{1-x}Mn_xTe$ system, it has been found that spin-flip scattering times are independent of the strong spin-spin exchange interaction between the carriers and the magnetic ions, in strong contrast with calculations which show that the s-d exchange is a very efficient spin-flip scatterer for electrons in these systems,[75] in the $Zn_{1-x}Mn_xSe$ system[101] the experiments are in agreement with the theoretical predictions. Time-resolved Kerr reflectivity experiments in modulation doped $Zn_{1-x}Cd_xSe$ QW's have shown that the electron spin polarization, at low temperatures, remain nearly three orders of magnitude longer than in insulating samples.[104, 105] In the same system, but undoped, pump-probe and degenerate four wave-mixing experiments have manifested that the decay of exciton spin, after resonant excitation, is faster than the exciton dephasing time, and this effect has been attributed to alloy disorder.[60]

Kuzma et al. have studied the spin polarization of a two-dimensional electron gas (2DEG) buried within a semiconductor heterostructure in the fractional quantum Hall regime.[71] Using a site-selective nuclear magnetic resonance technique they have discovered several unexpected properties of this system when the occupation of electrons is the fractional quantum Hall effect with $\nu=1/3$ Their data demonstrate that the spin polarization of the electron gas decreases as the system is tuned away from $\nu = 1/3$, revealing the presence of spin-reversal charged excitations. These findings suggest a remarkable decoupling between the energy of the two-dimensional electron gas spins and their environment. This is an important point for the fabrication of spin-polarized devices. A second important quality is the ability to manipulate the spin system, and Kuzma et al. have also shown that radio frequency radiation couples to these spin excitations. This suggests the exciting possibility that resonance techniques conventionally targeted at nuclear spins may ultimately prevail in controlling these electronic spins as well.

The studies of the dynamics of coherent control of excitons,[106-110] which provide a deep insight into interaction processes of excited states, have been also used to investigate the optical orientation of excitons in $Zn_{1-x}Mn_xSe$[97-100] and GaAs QW's.[57, 70, 111-114] In the work of the Toulouse group,[112] the optical dephasing time of excitons, their longitudinal and transverse spin relaxation times, and their radiative lifetimes, are measured with the same experimental conditions based on the linear response of the samples.

In this work, we will focus primarily on time-resolved polarized photoluminescence. The rest of the manuscript is organized as follows: Section 2 gives the experimental details. The spin-flip processes in intrinsic QW's are presented in Section 3.a. The splitting between polarized excitons is shown in Section 3b. Section 3c presents the spin decay and population distribution of an optically pumped 2DEG. The temperature dependence of the electron spin-flip times is given in Section 3d. Finally, we summarize in Section 4.

## 2. Experimental details

The experiments were performed in a temperature variable, cold finger cryostat exciting the samples with light pulses. These were obtained from either a mode-locked Nd-YAG laser, which synchronously pumped a cavity-matched dye laser, or a Ti:Za mode-locked laser pumped by an $Ar^+$-ion laser. The pulse width were 5ps and 1.2 ps for the dye and the Ti:Za systems, respectively. The incident light was directed along the growth axis of the heterostructures and a back-scattering geometry was used. The PL was time-resolved in a standard up-conversion spectrometer. The time resolution, obtained by overlapping on a non-linear crystal, $LiIO_3$, the luminescence from the sample with a delayed pulse from the laser, is basically determined by the temporal pulse width. A double grating monochromator was used to disperse the up-converted signal. The exciting light was circularly polarized by means of a λ/4 plate, and the PL was analyzed into its $\sigma^+$ and $\sigma^-$ components using a second λ/4 plate before the non-linear crystal. Time delays at a given emission energy or time resolved PL spectra at different delays after the excitation pulse have been obtained using this system.

For the spin relaxation of excitons we concentrate on the results of a GaAs/AlAs multiquantum well, consisting of 50 periods of nominally 77 Å-wide GaAs wells and 72 Å-wide AlAs barriers. The sample was exceptionally bright and presented a small Stokes shift of ~2.5 meV, which allowed us to perform quasi-resonant excitation at the free heavy-hole exciton, observed in pseudo-absorption experiments (PL excitation), detecting at the weakly-bound exciton seen in PL. The influence of the exciton localization has been reported in the literature.[55]

For the measurements of the electron spin-relaxation, we have studied four p-type modulation doped GaAs/GaAlAs quantum wells with hole sheet concentrations of ~3·$10^{11}$ $cm^{-2}$, mobilities of ~ 4000 $cm^2V^{-1}s^{-1}$ and well widths from 30 to 80 Å. The quantum well structures were grown either on [311]-GaAs substrates and modulation doped with Silicon or on [100]-GaAs substrates and doped with Beryllium. These structures were initially extensively tested with conventional, low power, cw experiments.[115] In this work, we concentrate on the results obtained in a 30 Å-thick QW modulation-doped with Beryllium. We have measured PL spectra at fixed times, as well as the PL decay at fixed emission energies, both as a function of temperature (T=10-50K) and as a function of the laser power, giving concentrations of excess carriers in the range between ~$10^{10}cm^{-2}$ and ~$10^{11}cm^{-2}$.

## 3. Results and discussion

### 3.a. Spin-flip processes of excitons and dark states

In this work, the spin for excitons and /or holes is defined as the third component of the total angular momentum (for electrons in III-V semiconductors, this coincide with the usual concept of spin since their orbital angular momentum is zer0, "s"-states). The spin relaxation of excitons between optically active states, with spin ±1, can take place in a single step, driven by the exchange interaction, or via a two-step process in which the individual constituents of the exciton, electron and hole, flip separately their spin, involving an intermediate dark state, with spin ±2. These paths are schematically depicted in Fig. 1. The rate of exciton-spin relaxation in the latter, indirect channel is limited by the slower single-particle spin-flip rate, which usually is the electron one.

Recent theoretical work on the spin relaxation of exciton-bound electrons shows that the coupling between optical active and inactive exciton states that differ only with respect to the electron spin direction represents an effective magnetic field that changes randomly as the exciton is elastically scattered an relaxes its spin.[87] In fact this mechanism leads to a very similar process to that responsible of the DP mechanism.[29]

In Figure 2, we show the cw-PL (points) and the photoluminescence excitation (PLE) spectra (line) of the studied sample. The spectra were recorded at 2K under very low excitation density. The Stoke shift between the *hh* exciton peaks in PL and PLE amounts to 2.5 meV. Increasing the temperature the Stokes shift decreases, and it vanishes at 40K. The linear temperature dependence of the decay time, which is characteristic of free excitons,[116] also indicates that localization effects are not very important. The upper panel shows the initial value of the degree of polarization of the emission, which for a given helicity of the exciting light, i.e., $\sigma^+$ is defined as the fractional difference of the PL intensities of the two circular polarizations, $\sigma^+$ and $\sigma^-$, $P=(I^+ - I^-)/(I^+ + I^-)$. These values were obtained at 5K from TR-PL experiments exciting with $\sigma^+$-polarized light and analyzing the two components of the emission. Exciting below the light-hole (*lh*) exciton, *P* presents a marked dependence on energy, increasing as the excitation energy decreases. At the energy of the *lh* exciton, *P* becomes negative an reaches a value of ~-20%. This negative value is due to excitonic effects, which enhance the creation of light holes and electrons with spin $+\frac{1}{2}$.

The time evolution of the PL for non-resonant excitation conditions is shown in Fig.3. The dark points correspond to the emission of +1 excitons (same polarization as the exciting light), while the open correspond to the annihilation of -1 excitons. Two main aspects are worthwhile to mention: in first place, the depolarization of the excitons occurs during the rising of the PL, while the initially created hot excitons are cooling down towards the lattice temperature. Secondly a very fast initial decay observed in the black $\sigma^+$-trace.

Several effects can give a fast initial decay of the exciton luminescence that are not related with the intrinsic radiative lifetime. Fast decays in the photoluminescence signal have been observed under resonant excitation conditions and have attributed to: a) very short radiative exciton lifetime;[117-119] b) scattering from **K**=0 optically active excitons to **K**>0, non-optically active, excitons[50, 118] and c) relaxation of the spin of one of the fermions of the exciton to relax the total spin of the exciton from ±1 to ±2.[50, 56, 58, 59, 94, 118, 120]. Excitation under non-resonant conditions rules out the first two explanations.[56] Therefore, only the transfer from optically active excitons to dark excitons can explain such as fast decay. It is worthwhile to mention that dark states have been observed directly in two-photon absorption measurements.[121]

In the detailed study by Vinattieri et al. is demonstrated that in spite of the large number of parameters needed to model this decay reliable values can be extracted from a careful fit procedure of the experimental curves.[120] The lines in Fig.3 correspond to the best fit to a dynamical model based on the scheme on the upper part of the figure. It is assumed that an initial population of +1 excitons are created in the upper state (with **K**>0), these have to relax ($\tau_k$) to the ground lower-state (**K**=0), before they recombine ($\tau_r$). During this time the excitons can flip the spin in a single step ($\tau_x$), or the electrons (holes) can flip their spins individually, $\tau_e(\tau_h)$. These spin-flip processes can happen in the ground and in the excited upper-sates. The energy separation between the dark, ±2, states and the active ones, $\Delta$ is given by the exchange interaction.

We find a value of Δ=80 μeV, in agreement with values reported in the literature,[113, 122-124] but significantly smaller than the 200μeV found in Ref. XX[58]. From our fitting, we also deduce that the active states lie above the dark states, in agreement with most of the other reports and in contradiction with the indications found in the time-resolved two-photon experiments.[58]

Our value for $\tau_h$ = 25±8 ps is compatible with others found in the literature,[45] especially when the hole belongs to an exciton;[47] although a much longer value of 1 ns has also been reported,[66] and Vinatteri *et al.* report values of $\tau_h$ =150-100 ps.[120] Similarly, the value of $\tau_e$ =320±100 ps is similar with the value of 250 ps predicted theoretically for a 80Å QW,[87] and in agreement with the values found in experiments in the presence of an external electric field (from 333ps to 3.3 ns).[120] Finally our fit obtains $\tau_k$ =80± 15 ps and $\tau_x$ =50±10 ps. The latter value is in good agreement with the results of Ref. 120, however our value of $\tau_k$ is significantly higher than the 18 ps, previously reported in this reference, as can be expected from the fact that our excitation conditions are not in resonance with the *hh* exciton.

The thick (thin) lines in Fig. 3 correspond to the active (dark) excitonic populations (only the former can be observed in a one-photon emission experiment). The large population of -2 states, manifests the importance of the rapid spin-flip of the holes which form the excitons. The inset in the figure depict these populations for the **K**>0, upper, excitons. Our results demonstrate that the transfer from optically active excitons (±1) to dark ones (±2) occurs also under excitation with circularly polarized light, in disagreement with the results of the group of Toulouse,[56, 59] who find this effect only when the exciting light is elliptically polarized.

All these facts indicate that exciton dynamics is very complex and is influenced by exciton recombination and momentum scattering and also by exciton, electron and hole spin dynamics.

### 3.b. Spin splitting in a polarized exciton gas

Figure 4 depicts TR-PL spectra taken at 8K 6 ps after excitation with $\sigma^+$-pulses for a density of 6.5x10$^{10}$ cm$^{-2}$, exciting at 1.625 eV. The gray area correspond to the polarized ($\sigma^+$, spin +1) emission while the dashed one shows the unpolarized ($\sigma^-$, spin –1) PL. A clear energy splitting of ~4.5 meV is clearly seen between the two peaks.

Increasing the excitation density, both a broadening of the lines and a strong enhancement of the splitting is obtained as shown in Fig. 5. The splitting is mostly due to the red shift of the $\sigma^-$ polarized emission and exhibits marked time and excitation-energy dependences.[93] It has been contradictorily reported that the splitting is either due to the energy shift of the luminescence component with the same helicity of the laser pump, the other component being only slightly red shifted,[91] or due to the component of the opposite helicity[92]. Theory predict that the absolute positions of the $\sigma^+$ and $\sigma^-$ emission components depends on the quasi-3D vs. quasi-2D character of the semiconductor system;[96] and recent unpublished experiments in double QW's demonstrate that those positions can be varied by an external electric field applied to the heterostructures.[125] One should also mention that in this case, where the excitons become indirect in real space with drastically reduced electron-hole overlap, the spin flip processes could be dominated by those of the electrons forming the excitons.[87]

Figure 6a depicts the dependence of the energy positions of the PL on the initial carrier density (open and solid points). Under the conditions presented in this figure,

12ps after excitation at 1.631eV, the $\sigma^+$ emission remains practically constant, while the $\sigma^-$ red shifts with increasing carrier density up to $9\times10^{10}$ cm$^{-2}$. The lines correspond to a model, which takes into account interexcitonic exchange interaction and screening,[96] that gives the changes in the energies of the interacting ±1 excitons as a function of the total and the ±1 populations of excitons as:

$$E^{\pm}(eV) = 2.214 x 10^{-16} \, a(Å) \times \left[1.515 n^{\pm}(cm^{-2}) - 0.41\pi n(cm^{-2})\right] \quad (1)$$

For the lines in Fig.6a, the energy of a single exciton has been taken from the experimental energy of the +1 exciton at the lowest carrier density; a three-dimensional Bohr radius, $a$, of 150Å and an initial degree of polarization of 80% have been used. In spite of the strong approximations used in the theory, such as neglecting the presence of dark, ±2, states and assuming that the excitons are all at **K**=0, which are not borne out by the experiments, the agreement with the experiments is satisfactory.

The splitting is strongly correlated with the degree of polarization of the exciton gas, as can be observed in Fig. 6b, which depicts the time evolution of the splitting and $P$. This has also been confirmed by experiments where the degree of polarization of the exciting light has been varied from circular to linear.[92] This correlation is also predicted by the theory: Eq. (1) displays that the splitting is proportional to the difference between exciton +1 ($n^+$) and -1 ($n^-$) populations and thus proportional to the degree of polarization [$P=(n^+ - n^-)/(n^+ + n^-)$]. When the $n^+$ and $n^-$ populations become similar, and therefore $P$ approaches zero, the splitting disappears as a consequence of the convergence of the ±1 excitons towards the same energy, as predicted by Eq. (1). The time decay of $P$ originates from the excitonic spin-flip processes described in Sect. 3a, which are mainly driven by intraexcitonic exchange interaction (Bir-Aronov-Pikus, BAP, mechanism).[45]

Additionally, it has been also observed under resonant excitation that the luminescence component co-polarized with the laser excitation is narrower than the counter-polarized one, despite the much higher density of +1 excitons.[70] The linewidth difference decreases with time due to the decay of the PL polarization.

### 3.c. Spin polarization of an optically pumped electron gas

The time evolution of the two spin components of a photocreated 2DEG is investigated as a function of the density of carriers excited with a picosecond laser pulse. Exciting a p-type GaAs quantum well below the light-hole resonance, electrons with almost purely one spin component are photocreated. Filling of the conduction band is clearly different for both electron-spin components, leading to an appreciable shift between $\sigma^+$ and $\sigma^-$ emission spectra. The decay of spin polarization of the electron gas is found to depend strongly on the excitation power: an usual monoexponential decay of spin polarization (with a characteristic time of 550 ps) is observed at low powers; whereas a fast depolarization process (characteristic time of 20 ps) turns on progressively when the density of photocreated carriers approaches the concentration of holes originating from doping. The observation of the fast component (typical for the relaxation of hole magnetic moment) in the electron-spin relaxation suggests that this is driven by the decay of the total polarization of the hole gas. Such process may only be expected at sufficiently high excitation powers when concentration of photocreated, spin-polarized holes becomes comparable with the density of non-polarized holes arising from doping.

Figure 7 depicts the PLE spectra of a 30Å, [100]-oriented QW at 2K, exciting with $\sigma^+$-polarized light recorded at the tail of the PL. The onset of the absorption due to the *hh* transition is clearly seen as a peak in the $\sigma^+\sigma^+$ spectrum at 1.7 eV, while the one corresponding to the *lh* transition is dominant in the $\sigma^+\sigma^-$ spectrum at 1.738 eV. As can be deduced from this figure, the sample shows a high degree of optical alignment, indicating a long spin-flip relaxation time for the photoexcited electrons. This is even more clearly seen in the inset of the figure, which displays the cw-PL of a 50Å-wide, [100] QW at different lattice temperatures. The thick (thin) lines correspond to the co-polarized (counter-polarized) emission with the exciting light. A very large degree of polarization is observed at the lowest T and it diminishes as the temperature is increased. For excitation energy above the threshold of *lh* states, the polarization decreases rapidly, since electrons with opposite spin direction are produced from the light-hole subband. Large values of *P* have been already reported in p-doped strained films,[126] quantum wells[127] and superlattices.[128]

An extremely long electron-spin-relaxation time of 20 ns, two orders of magnitude longer than that found in homogeneously doped GaAs for comparable acceptor concentrations, has been found in p-type δ-doped GaAs:Be/AlGaAs double heterostructures and has been attributed to a drastic reduction of the electron-hole wavefunction overlap, which strongly reduces the electron-hole exchange interaction.[67] In II-VI based QW's, $Zn_{1-x}Cd_xSe/ZnSe$, the introduction of carriers by modulation doping increases the electronic spin lifetimes several orders of magnitude relative to insulating counterparts, a trend that is also observed in bulk semiconductors.[104, 105] The spin lifetime exceeds the recombination time by nearly two orders of magnitude, suggesting that the 2DEG acquires a net polarization either through energy relaxation of spin-polarized electrons or through angular-momentum transfer within the electronic system. These studies also show that the nanosecond spin-flip times last up to room temperature. Also for excitons, spin-flip times exceeding the radiative recombination lifetime have been observed in InGaAs quantum disks.[62]

Using these p-type samples we can easily investigate spin alignment effects in the conduction band: electrons with an unbalanced population of the two spin components, created under circularly polarized excitation, recombine with non polarized holes which mostly originate from doping. Therefore the spin relaxation of the electrons can be obtained from the difference of the time evolution of the two orthogonally polarized emissions. The spin-relaxation time of photogenerated electrons in p-doped QW's has been calculated by Maialle and Degani (Ref. 79).They have found that, for the mechanism of exchange interaction, the spin mixing of the valence hole is not important due to a compensation between the enlargement of the hole density of states and a spin-mixing induced decrease of the exchange strength.

In an ideal sample, using the estimated hole Fermi energy of 2 meV, and assuming a ratio of six between the hole and electron effective masses, the circularly polarized luminescence spectra are expected to directly reflect the distributions of electrons with different spin components in an energy range up to ~12meV above the conduction-band edge. However, as shown in Fig.8, the spectra measured at 15 ps after a 20 mW excitation cover a somehow wider spectral range. We assume that even these spectra probe only the properties of the electron gas since the investigated sample shows an appreciable spectral broadening already under low-power cw excitation (half width of the luminescence ~10 meV).

Figure 8 represents the $\sigma^+$ and $\sigma^-$ components of the luminescence spectra excited with $\sigma^+$ pulses at 1.717 eV. These spectra clearly show the difference in the occupation of electronic states with opposite spins. This difference, which vanishes at longer delay times, leads to the energy difference between the positions of the maxima in the $\sigma^+$ and $\sigma^-$ luminescence spectra.

The most conspicuous finding is illustrated in Fig. 9 : panel a) shows the evolution of the two components of the polarized PL and panels b) and c) depict the time evolution of the polarization degree, *P*, for two different powers of the exciting light. We have found that the decay of spin polarization of the electron gas depends very much on the intensity of the laser excitation. The decay of the degree of luminescence polarization, *P*, measured at T = 8 K can be well reproduced by a sum of two exponential decays with two distinct characteristic times of $\tau_1$= 20 ps and $\tau_2$= 550 ps. The amplitude of the fast component vanishes at low excitation power, whereas this fast process almost completely determines the electron spin depolarization at the highest level of laser excitation. It is well known that for any mechanism of electron spin relaxation,[29] the probability of spin flip transitions increases as a function of the electron k-vector. Therefore, the increase in the rate of spin relaxation as a function of the laser power may result partially from a larger electron kinetic energy caused either by an increase of the effective electron temperature or of the initial electron concentration. However, this simple reasoning hardly explains our data. From an analysis of the time evolution of the luminescence spectra, we have estimated an increase in the mean kinetic energy of the electron gas of only 4 meV at the highest laser power. This amount is not sufficient to reduce the spin relaxation time down to 20 ps, since at low excitation powers but high lattice temperatures ( kT ~ 3.44 meV ) we still observe a relatively long spin relaxation time ( ~ 80 ps ).

We have simulated measured spectra at different times after the excitation, $I^{+(-)}(\hbar\omega)$, by the broadened convolution of Fermi-Dirac statistics for non-polarized gas of holes and two spin components of the 2DEG, assuming the conservation of **k**-selection rules.

$$I^{+(-)}(\hbar\omega - E_g) = I^{+(-)}(E_e + E_h) = A \int_0^\infty f_{E_e'}^{+(-)} f_{E_h} G_\Gamma(E_e - E_e')\delta(\vec{k}_e' - \vec{k}_h')dE_e' \quad (2)$$

Here $E_g$ is the energy gap; e(h) stands for electrons(holes); $I^{+(-)}$} denotes the intensity of $\sigma^+$ ($\sigma^-$) luminescence, $G_\Gamma(x)$ is a Gaussian broadening function with a broadening parameter, $\Gamma$ of 7 meV, chosen to reproduce the low-temperature (4 K), low-power (1 mW/cm$^2$ cw-spectra, $E_{e(h)}=\hbar k^2_{e(h)}/2m_{e(h)}$ is the electron (hole) energy, where we assumed $m_e$=0.075$m_0$ and $m_e/m_h$=0.18.

An analysis of pairs of $\sigma^+$ and $\sigma^-$ PL spectra leads us first to conclude that each component is well described assuming a common temperature for the two electron spin components (and for holes), but different values of the chemical potential. Excluding very short delay times after excitation, i.e., already after a few picoseconds, each spin component of the electron gas can be qualitatively characterized by its own Fermi distribution, each one with different chemical potential but both with very similar temperatures. It could be argued that due to differences in the exchange interaction, the unbalanced populations of the two spin components could also induce a difference in the many-body renormalization between occupied electronic states with opposite spins. We have systematically observed that the low frequency onset of the emission associated with the majority spins sets out always below the corresponding onset of the minority-spin luminescence. This effect is, however, rather weak as can be seen from the spectra shown in Fig. 8. This means that, under our experimental conditions, the exchange interaction between electrons is unable to stabilize a common chemical potential of the electron gas for the two spin components. This latter situation might be expected under equilibrium conditions and would imply a difference in the renormalization of the conduction band edges for the two spin components. On the other hand, the attainment of a common temperature for electrons and holes in a very short time is expected and it is well documented in the literature.[129]

To fit our measurements, we assumed that the time evolution of the total electron concentration follows the decay of the total luminescence intensity observed at sufficiently long delay times, i.e., $n = n^+ + n^- = N_0 \, e^{-t/230ps}$, in the case of the 10mW-series. The initial electron concentration $N_0 = 15 \times 10^{10}$ cm$^{-2}$ was found by self consistent fitting of several spectra measured at long delay times, and agrees within a factor of 2 with an estimation based on the absorption coefficient and the laser power density on the samples. Hole concentration was assumed to be $n_h = n^+ + n^- + n^0_h$ where $n^0_h = 3 \times 10^{11}$ cm$^{-2}$ originates from modulation doping. Electron concentrations $n^{+(-)}$, which define the corresponding chemical potentials, were determined from the experiment assuming $n^+/n^-$ to be equal to the ratio of the integrated intensities of the $\sigma^+$ and $\sigma^-$ PL. Finally, a given pair of $\sigma^+$ and $\sigma^-$ spectra was fit with only two parameters: carrier temperature and a proportionality factor, *A*, which was found to be common for all the simulated spectra, within experimental error.

The obtained time evolution of carrier temperature, is shown in Fig.10 (open circles). The carrier temperature rises up to ~100K just after the laser pulse. This fact, in conjunction with the measurements as a function of lattice temperature (see Section 3.d), accounts for the fast depolarization of electronic spins, induced by the laser power. Our results confirm the high efficiency of carrier-carrier interaction in establishing a common temperature for electrons and holes. Cold before excitation, the gas of holes becomes nondegenerate almost immediately after the laser pulse. This nondegenerate character of the hole gas is illustrated in Fig.10 (solid circles), where the number of occupied hole states at the top of the valence band is plotted as a function of time.

High carrier temperatures, and the associated fast depolarization of electronic spins, shortly after high-power pulsed excitation, are not very surprising though we show that the degree of spin polarization is a sensitive measure of carrier temperature. On the other hand, it is interesting to note that at long delay times (low carrier temperature), we always observe a slow spin relaxation, independently of the excitation power, i.e., electron concentration. From the spectral simulation we have, for example, concluded that for a 40mW-power excitation and 175 ps after the laser pulse, the carrier temperature is 15K and electron concentration is $2.5 \times 10^{11}$ cm$^{-2}$. Under these conditions both electrons and holes are degenerate and electrons flip the spin in the vicinity of their chemical potentials (which are slightly different for both spin up and spin down components). The electrons flipping the spin have high kinetic energies ($E_F/k = 90$K); however, the observed spin relaxation time remains slow. This is in contrast to the case of fast spin relaxation (short times) when the electrons flipping spin have also high kinetic energy (raised by temperature), but carrier distributions are more Boltzmann-like.

Bir *et al.* have established that the rate of electron spin relaxation due to holes is proportional to the time of interaction with the holes, i.e., the time during which the distance between them is less than the electron wavelength.[32] Under conditions when this time equals the time of diffusion of holes through the interaction region, strong scattering of holes lead to a decrease in the electron spin relaxation time. On the other hand, under conditions when the hole spin relaxation time becomes less than the interaction time, strong hole spin relaxation leads to a decrease in the electron spin relaxation rate due to an efficient averaging of the hole spin. Therefore, the non-degenerate character of the electrons and holes, and their high temperatures, shortly after excitation can lead to the very rapid electron-spin relaxation found on our experiments.

We therefore conclude that fast spin depolarization in our structures is driven by the nondegenerate character of carrier distribution and not exclusively by the increase of the electron kinetic energy. Such behavior can be understood in terms of the BAP mechanism of the electron spin relaxation but it is hardly accounted for by the DP processes whose efficiency is directly related to the electron kinetic energy. As can be deduced from our previous discussion, nondegenerate carrier distributions favor the efficiency of spin-flip electron scattering via the exchange interaction with holes, in contrast, the available number of scattering configurations is appreciable reduced for the degenerate systems. Similar results could be also expected in n-doped samples, which have been shown recently to have long electronic spin lifetimes,[104] although the very fast spin relaxation of photocreated holes[45] would render the experiments much more difficult.

The splitting in the PL maxima is linked to the differences in the Fermi energies of spin-up and spin-down electrons. Its time-evolution and power dependence is strongly connected to the spin-flip dynamics. Figure 11 depicts the energy of the maxima of the PL for different excitation powers as a function of time delay after the pulsed excitation. At 10 mW, both the co-polarized (solid points) and the counter-polarized (open points) red shift with time and the splitting vanishes at ~400 ps. Doubling the power a distinct behavior is found for the $\sigma^-$ emission, which blue shifts during the first ~75 ps; the splitting disappears now at ~250 ps. This time is further reduced to ~75 ps when the power is again augmented by a factor of two, and the blue shift of $\sigma^-$ last until the energy positions of the two polarizations merge. Additionally, it is observed in the figure that the initial splitting, at t~0, increases from ~2meV (10 mW) to ~6meV (40 mW). The inset in the figure gives a qualitative description of the splitting. Initially a larger population of +½ electrons is created by the $\sigma^+$ pulse compared to that of -½ electrons, and therefore $E_F(+½) > E_F(-½)$. The recombination processes (curved-solid arrows) lower the populations of both kinds of electrons, thus also decreases their Fermi energies and a red-shift of both emissions results. However, the spin-flip processes (white arrow) decreases (increases) the populations of +½ (-½) electrons. Depending on which process, recombination vs. spin-flip, is faster, a red or blue shift of the $\sigma^-$ emission is obtained, while the $\sigma^+$ PL will always red-shift. Increasing the power the fast channel for spin-flip (20 ps) becomes more important, and since it is much faster than the recombination (230 ps) a blue shift is obtained. In contrast, at low powers the spin-flip becomes slower (550 ps) than the recombination and both co- and counter-polarized emissions red-shift.

This finding of two different quasi-Fermi levels could have important consequences for possible ultrafast devices based on spin-polarized transport in hybrid magnetic-semiconductor systems. Transport experiments through nonmagnetic metal sandwiched between two ferromagnetic films have shown that is valid to assume that the current is carried by two non-intermixing components, spin up and down, and that one need only to determine the spin scattering coefficient for each of these components to completely describe the magnetoresistence behavior of a multilayered structure.[1] A spin transistor (emitter and collector: ferromagnetic films, base: non-magnetic metal), which is based on the shift of the chemical potential due to the accumulation of spin-polarized electrons in a normal metal, has been also demonstrated.[130]

### 3.d. Temperature dependence of the electron spin-flip

Figure 12 shows the PL dynamics for different lattice temperatures at a photocreated electron density of $2.6 \times 10^9$ cm$^{-2}$. Increasing T, a large increase of the decay time, $\tau_d$, is observed from 195 ps at 5 K to 475 ps at 60 K. This increase is attributed to heating of the carriers due to electron-phonon scattering, which moves the electrons out of the emission region, therefore competing with the radiative recombination processes. Furthermore, with increasing lattice T, the electron and hole distribution functions broaden yielding a smaller density of occupation and thus decreasing the PL intensity as can be readily seen in Fig.12. The rate of increase of $\tau_d$ is independent of the growth direction, but increases considerably with increasing QW width; for 80Å QW attaining $\tau_d$=780 ps at 60 K.

The time evolution of the degree of polarization, *P*, at 8K for the [100] QW, exciting at 1.717 eV and detecting at the PL peak, is shown in Fig. 13a for a photocreated electron density of $5.3 \times 10^9$ cm$^{-2}$. It is clearly seen that the decay is not monoexponential, indicating the presence of different spin-flip mechanisms. A fit with the sum of two exponential decays (lines in Fig. 13a) obtains spin-flip times of $\tau_1$=20 ps and $\tau_2$=550 ps. The fast time, $\tau_1$, is attributed to the nondegenarate character of the holes almost immediately after the laser pulse. This carrier distributions was shown in the previous Section to favor the efficiency of spin-flip electron scattering via the exchange interaction with holes. The slow time, $\tau_2$, corresponds to the spin flip of electrons in the presence of a degenerate hole gas; it is important to note that $\tau_2$ is a factor of two longer than the PL decay time, $\tau_d$. Both times are independent of the excitation power, but the contribution to the spin relaxation of the fast mechanism increases with power. Increasing T, both spin-flip mechanisms speed up considerably. A fit at 40 K obtains values of 10 ps and 80 ps for $\tau_1$ and $\tau_2$, respectively. We will concentrate in the following on $\tau_2$, which we identify as the intrinsic spin-flip of electrons, $\tau_{sf}$.

In bulk GaAs the temperature dependence of the spin flip of electrons has been studied in great detail by Fishman and Lampel,[15] who found that at low temperature the exchange interaction with the holes is the dominant relaxation mechanism. On the other hand, at higher temperatures (T≥100K), and low acceptor concentrations, the DP mechanism governs the electron spin-relaxation. This mechanism is important only at high temperatures in 3D systems, because only then the thermal activation of the carriers impels them to feel the non-parabolicity of the conduction band.[17] However, in 2D systems, this mechanism should be more effective for thinner wells, where the quantum confinement effects are more marked. This has been observed in TR-PL measurements in p-doped GaAs QW's of different thickness:[68] for 180Å wide wells no clear dependence of the spin-flip time with temperature was found; however, for 55Å wells the DP mechanism was found to dominate for temperatures as low as 7K. It should be also mentioned that in spite of the large excitation powers used in this work, only a single, very long spin-flip time of the order of 1 ns at low temperatures is reported,[68] in contrast with the two channels that we find in our samples.

The T dependence of the spin-rate, $1/\tau_{sf}$, is depicted with solid points in Fig.13b, it allows the identification of the spin-flip mechanisms: for BAP a $T^{1/2}$ is expected,[32] while DP predicts a $T^\alpha$ dependence.[29, 75] The best fit of our data to those laws obtains $\alpha$=2.6±0.3, and proofs that, similarly to electrons in bulk, the spin-flip processes of two-dimensional electrons are governed by the exchange mechanism at low temperatures, while the DP mechanism takes over at higher temperatures. Furthermore, from our data a momentum scattering rate with a ~$T^{-1/2}$ dependence is deduced. This value of $\alpha$ is also compatible with the fact that the spin-flip of electrons takes place in a region of strong hole spin relaxation.[32] Finally, let us mention that the temperature dependence of the excitonic spin-flip is still not well understood in intrinsic QW's: although exchange interaction (BAP) is believed to be the main spin-flip mechanism, it is found that $\tau_{sf}$ is independent of T at low temperatures (4K<T<30K).[55, 58]

## 4. Summary

We have shown that spin-flip processes in 2D systems constitute a very active field of research and that still many questions are open to further investigations. In particular, we have shown that for excitons the main spin-flip mechanism is intraexcitonic exchange (BAP), although spin-flip of the individual fermions forming the exciton is responsible for fast decays observed in the TR-PL emission. Interexcitonic exchange is shown to be responsible for a splitting of the excitons in a polarized gas. In the case of a 2DEG the spin-flip of electrons is driven by the BAP and DP mechanisms at low and high temperatures, respectively. When this gas is created polarized by means of optical orientation and becomes dense, strong effects, nonlinear in the excitation power, are observed in the polarization the emission, which originate from the nondegenerate character of the carrier distribution at short times after the laser excitation. An optically aligned, spin-polarized electron gas can be well described by two separate Fermi-Dirac distribution functions, one for each spin component, with common temperature but different chemical potentials.

**Acknowledgments**

This work would not have been possible without the agreeable collaboration with many people, especially with: L. Muñoz, E. Pérez, J. Fernández-Rossier and C. Tejedor, who made a great work in the study of excitons. The studies of 2DEG were motivated by a collaboration with M. Potemski and L. Gravier; M.D. Martín contributed considerably to these studies. All the samples mentioned in this manuscript were kindly provided by K. Ploog. This research has been partially supported by the Fundación Ramón Areces, the Spanish DGICYT under contract PB96-0085 and the CAM (contract 07N/0026/1998).

**Figure Captions**

**Figure 1.-** Schematic representation of excitonic spin-flip. An optically active exciton ±1 can flip its spin to ∓1 in a single step (vertical path) or by sequentially flip of the spin of its fermions : first hole and then electron going from +1, passing through the dark state -2, to +1 (left path) or first electron and then hole, passing through the dark state +2 (right path)

**Figure 2.-** a) Photoluminescence (open points) and excitation spectra (line) of a 77 Å-wide GaAs recorded at 2K under an excitation of 5 mW cm$^{-2}$. The peaks are the heavy-hole (*hh*) and light-hole (*lh*) excitons. The step correspond to the *hh* subband continuum.. b) Initial degree of the polarization as a function of excitation energy obtained from time-resolved PL experiments at 5K.

**Figure 3.-** Time evolutions of the $\sigma^+$ (solid circles) and $\sigma^-$ (open circles) photoluminescence. The lines depict the best fits to a dynamical model based on the levels shown above the figure (see text). The inset represents the populations for the upper states ("u") obtained from the fits.

**Figure 4.-** Low temperature, 8K, time-resolved PL spectra of the sample shown in Fig.1 taken 6 ps after excitation with $\sigma^+$-polarized light at 1.625 eV. The gray (dashed) area depict the $\sigma^+$ ($\sigma^-$) emission. The initial carrier density is 6.5x10$^{10}$ cm$^{-2}$. The $\sigma^-$ emission has been enlarged by a factor of 3.5.

**Figure 5.-** Evolution of the time-resolved PL spectra taken 10 ps after excitation for the 77Å GaAs QW for different initial carrier densities. Each pair of spectra correspond to a given initial carrier density, those spectra lying at higher (lower) energy depict the co-polarized, $\sigma^+$ (counter-polarized, $\sigma^-$) PL. The arrows indicate the blue (red) shift of the $\sigma^+$ ($\sigma^-$) PL

**Figure 6.-** a) Energies of the co-polarized ($\sigma^+$, solid points) and counter-polarized ($\sigma^-$, open points) luminescence as a function of carrier density. The positions are taken 12 ps after excitation at 1.631 eV. The lines represent the results of Eq. (1). b) Time evolution of the PL splitting (solid points) and polarization (open points) for an initial carrier density of 5x10$^{10}$ cm$^{-2}$. The line shows the best fit to an exponential decay with a time constant of 41 ps.

**Figure 7.-** PL (solid line) and excitation spectra of a 30Å thick p-type modulation doped (p=3x10$^{11}$ cm$^{-2}$) QW measured for aligned (solid points) and crossed (open symbols) polarization of the exciting/emitted light. The shaded area indicates the energy used for excitation on the time-resolved experiments. The insets displays cw-PL spectra of a similar heterostructure, QW 50 Å thick, for different lattice temperatures and polarization configurations : aligned, thick lines and crossed, thin lines.

**Figure 8.-** Co-polarized (solid points) and counter-polarized (open points) components of the PL spectra of the sample shown in Fig. 7, measured at 15 ps after a $\sigma^+$-polarized pulsed excitation with a mean laser power of 20 mW and excitation energy of 1.717 eV. Bath temperature 8 K.

**Figure 9.-** a) Decay of $\sigma^+$ (solid points) and $\sigma^-$ (open points) luminescence under pulsed $\sigma^+$ excitation with a mean laser power of 1 mW. b) Time evolution of the polarization degree for 1 mW mean laser power. c) Same as b for a laser power of 40 mW. The lines represent the best fits according to the sum of two exponential decays : A x exp(-t/$\tau_1$)+B x exp(-t/$\tau_2$). The sample is the same as in Fig. 7.

**Figure 10.-** Time evolution of the carrier temperature (open circles), for a mean laser power of 10 mW, obtained from the fits of the time-resolved spectra using Eq. 2 (see text). The sample is the same as in Fig.7. The solid symbols depict the time evolution of the number of occupied hole states at the top of the valence band.

**Figure 11.-** Energy positions of the PL maxima for co-polarized (solid points) and counter-polarized (open symbols) emission as a function of the time delay after a pulsed $\sigma^+$ excitation for a mean laser power of : a) 10 mW ; b) 20 mW and c) 40 mW. The sample is the same as in Fig.7. The inset shows schematically the Fermi distributions, 2D density of states (DOS) and the corresponding electron +½ (dark area) and -½ (light area) populations, which decay either by recombination (thin, solid arrows) or spin-flip (thick, white arrow).

**Figure 12.-** PL time evolutions for different lattice temperatures of the 30Å-wide, modulation-doped QW shown in Fig. 7. The lines are best fits to exponential decays, with a decay time $\tau_d$.

**Figure 13.-** a) Time evolution of the PL polarization degree of the sample shown in Fig. 7. The lines represent the best fits according to the sum of two exponential decays, similarly to Fig. 9. b) Temperature dependence of the spin-flip rate of the slow channel, $\tau_2$, the lines are the fits to different spin-flip mechanisms (see text).

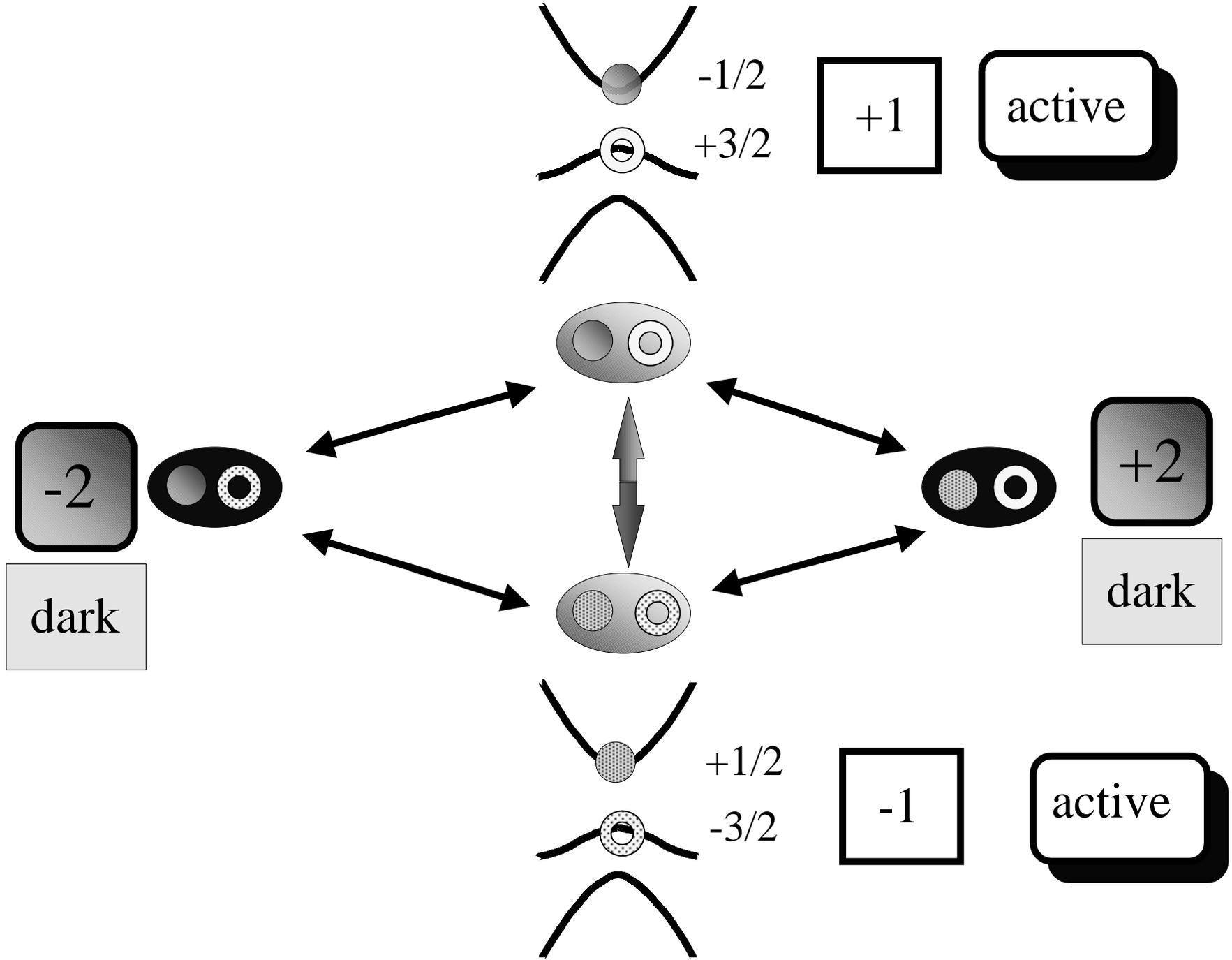

Fig.1 (Viña)

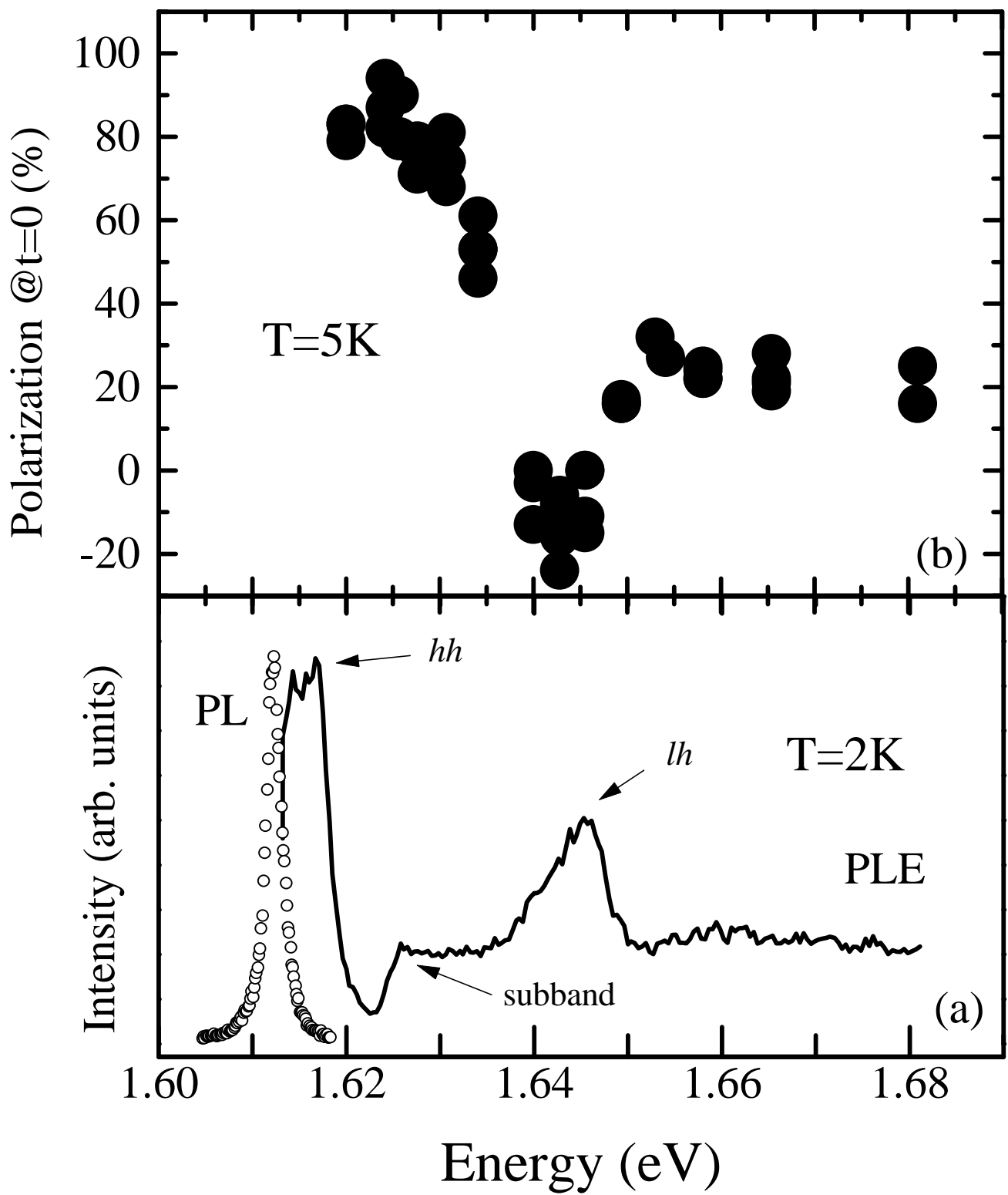

Fig.2 (Viña)

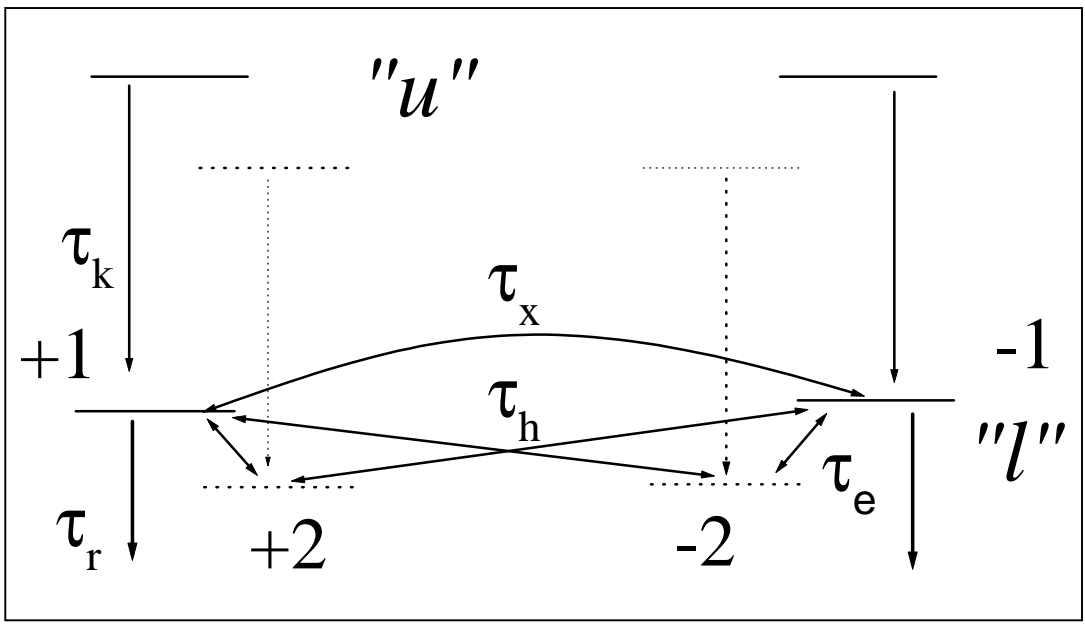
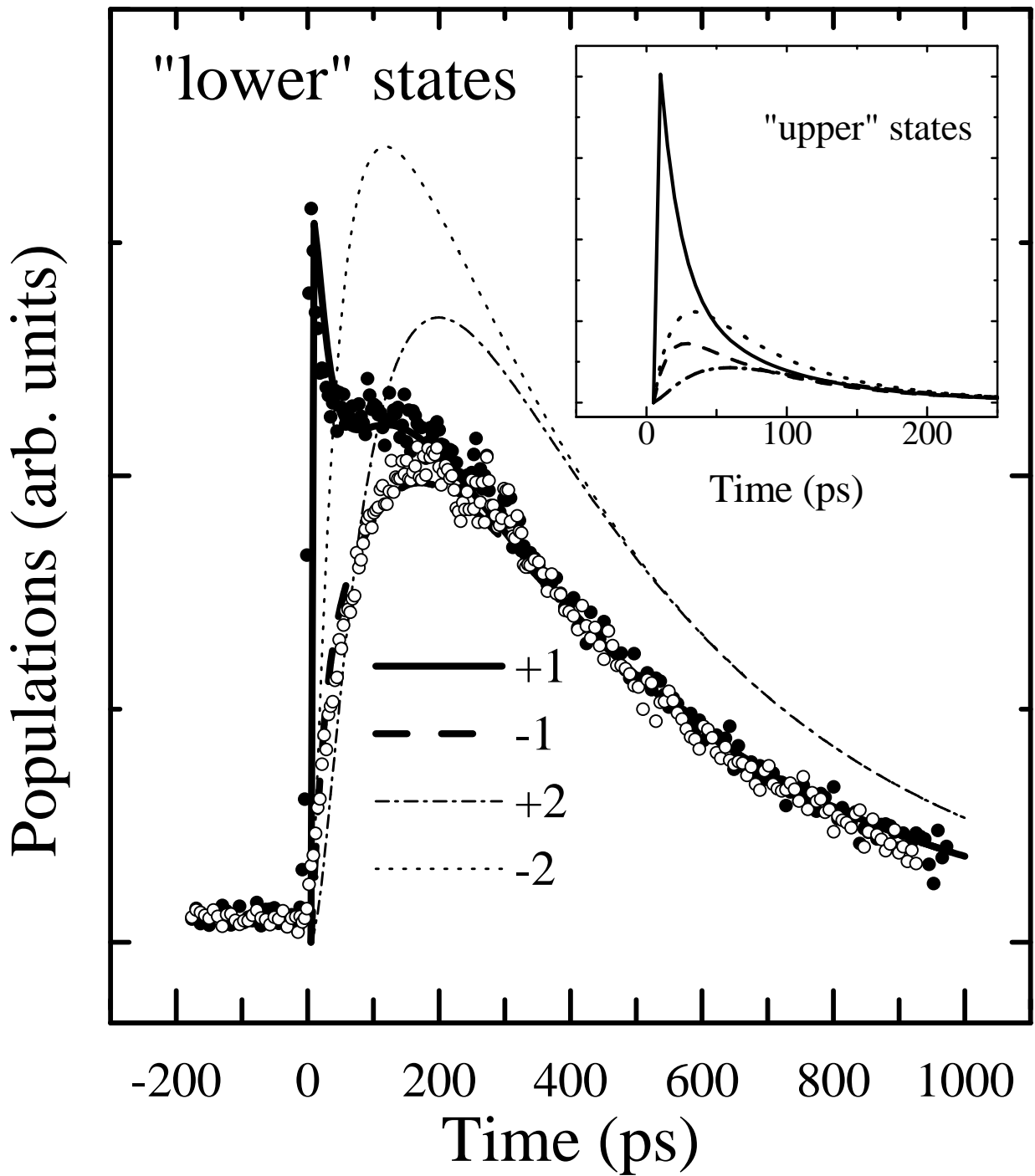

Fig.3 (Viña)

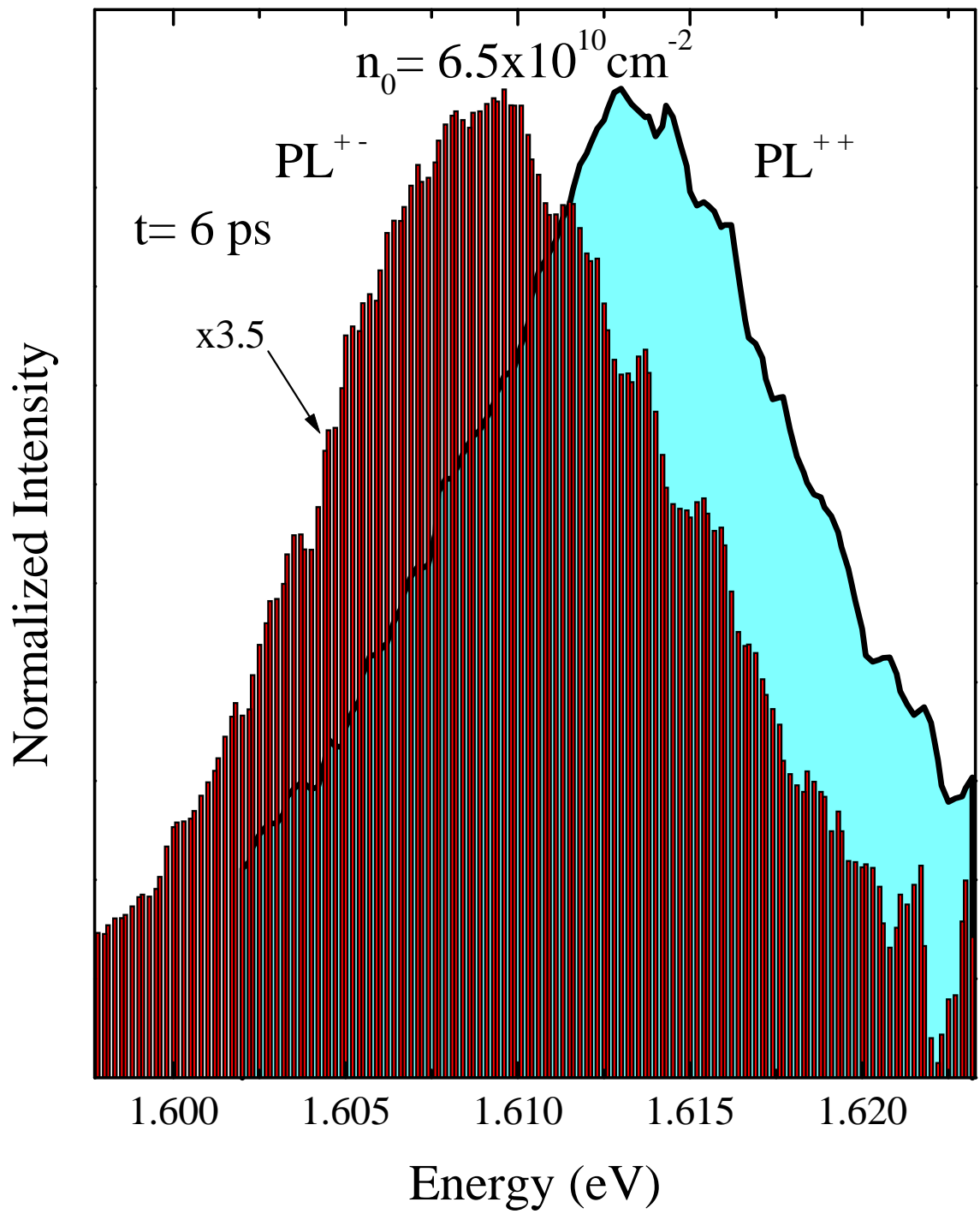

Fig. 4 (Viña)

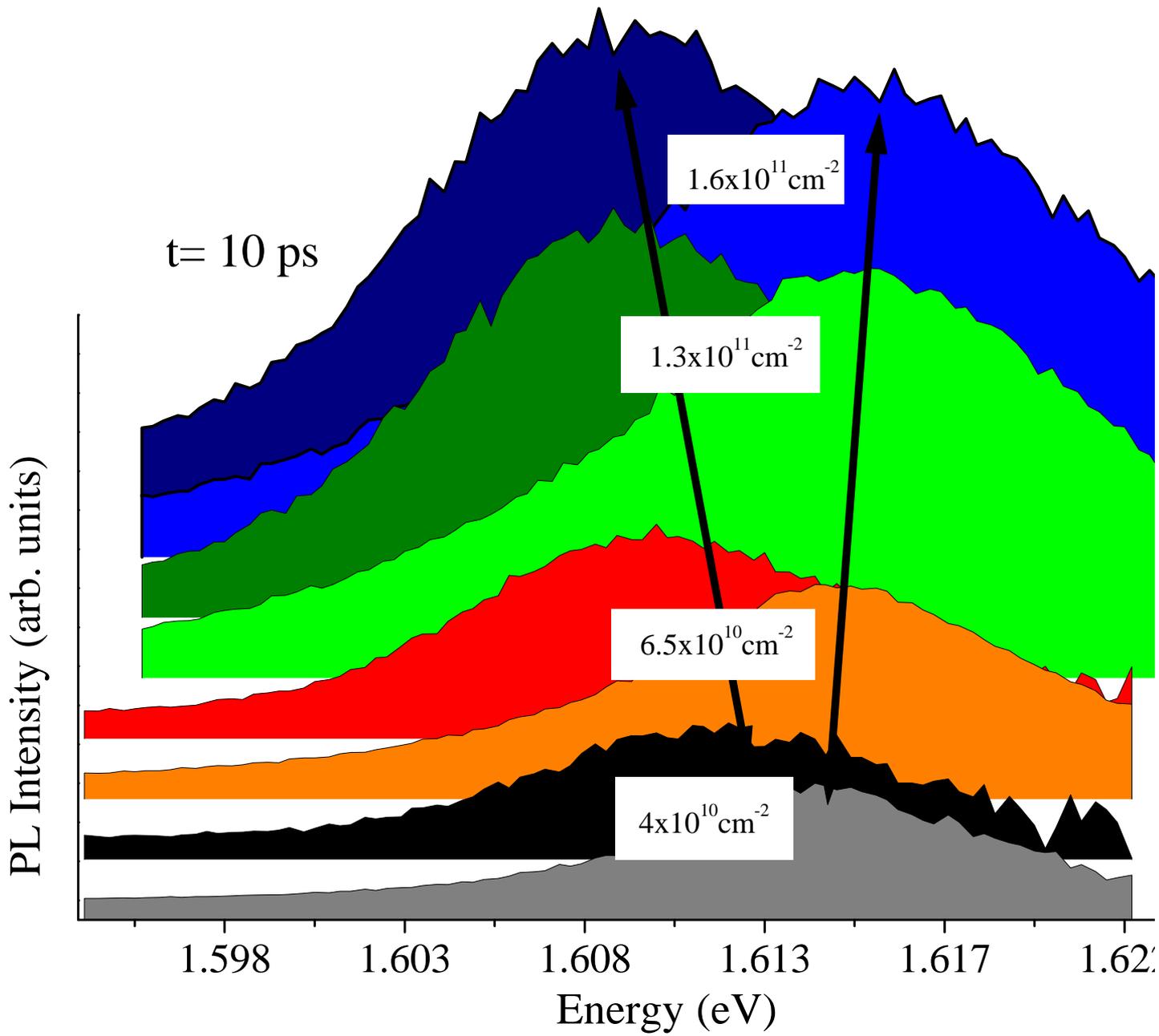

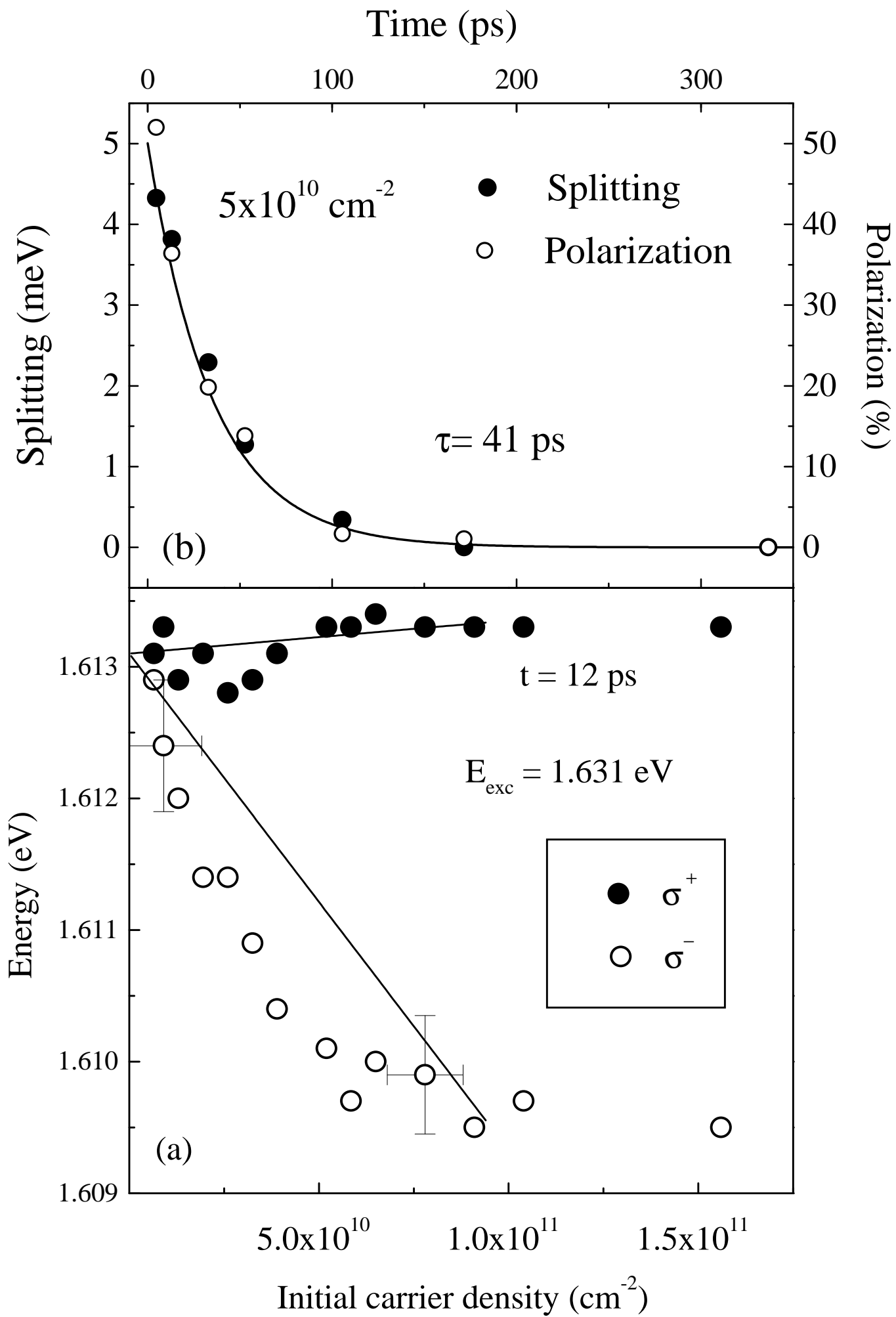

Fig. 6 (Viña)

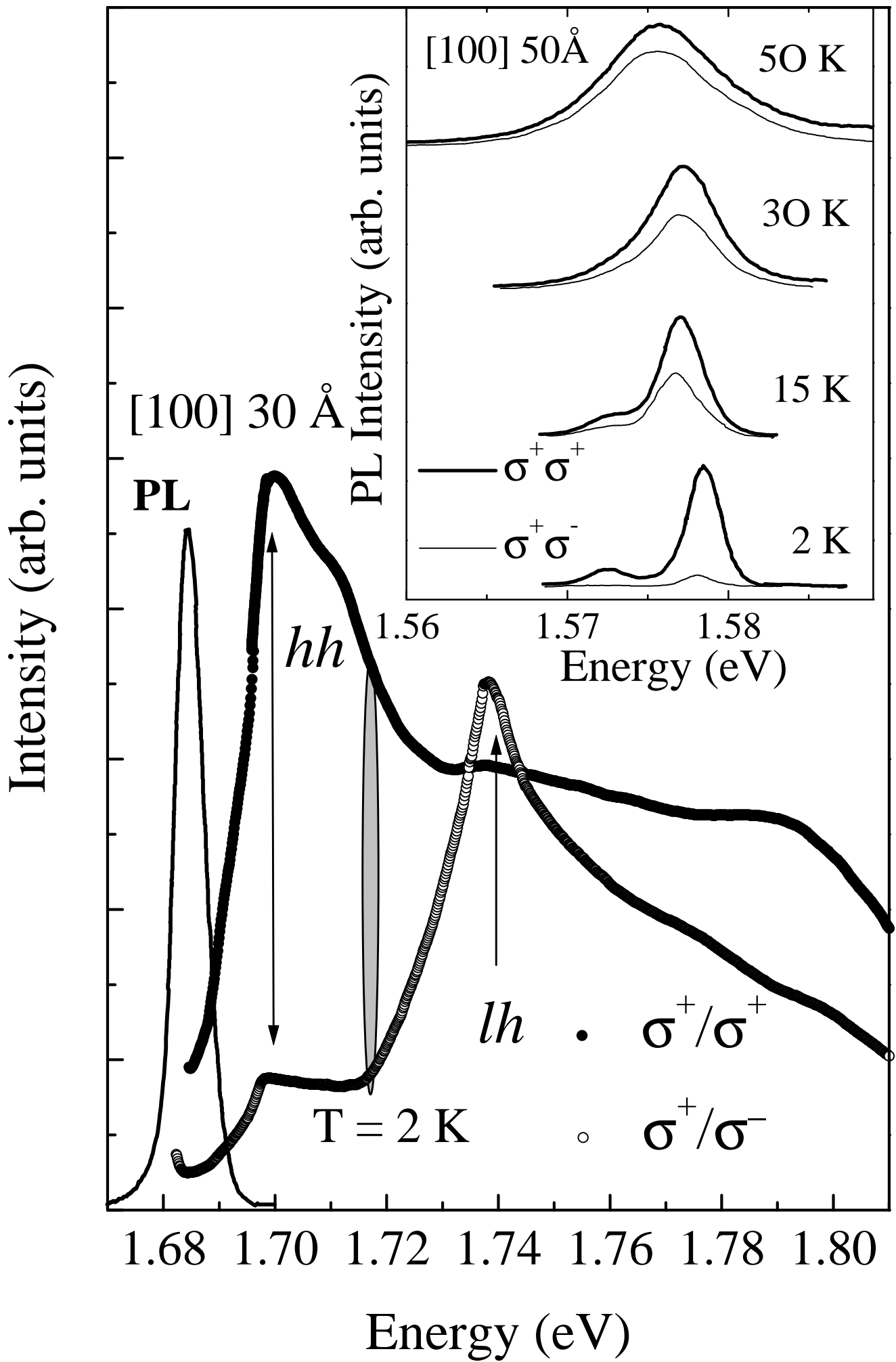

Fig. 7 (Viña)

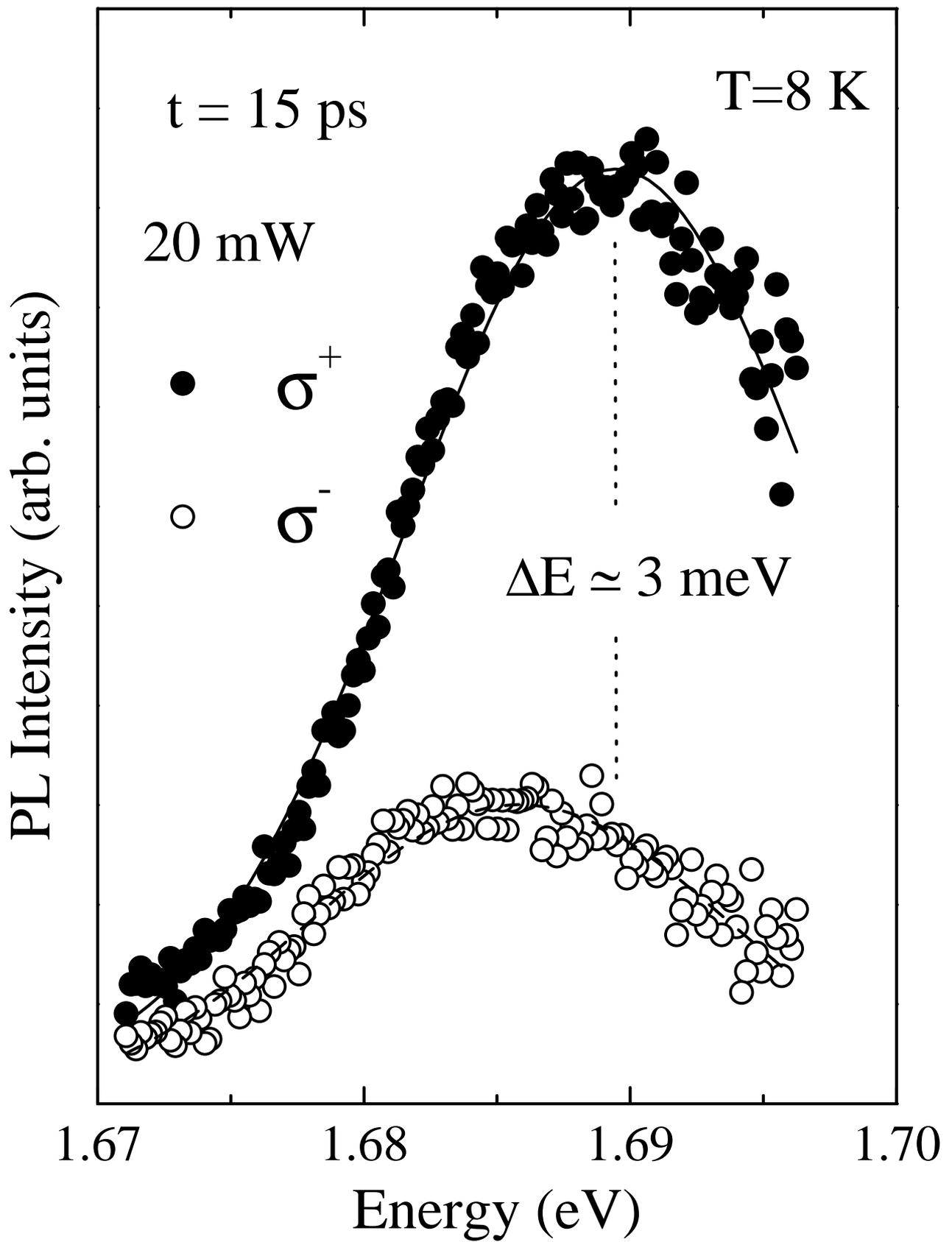

Fig. 8 (Viña)

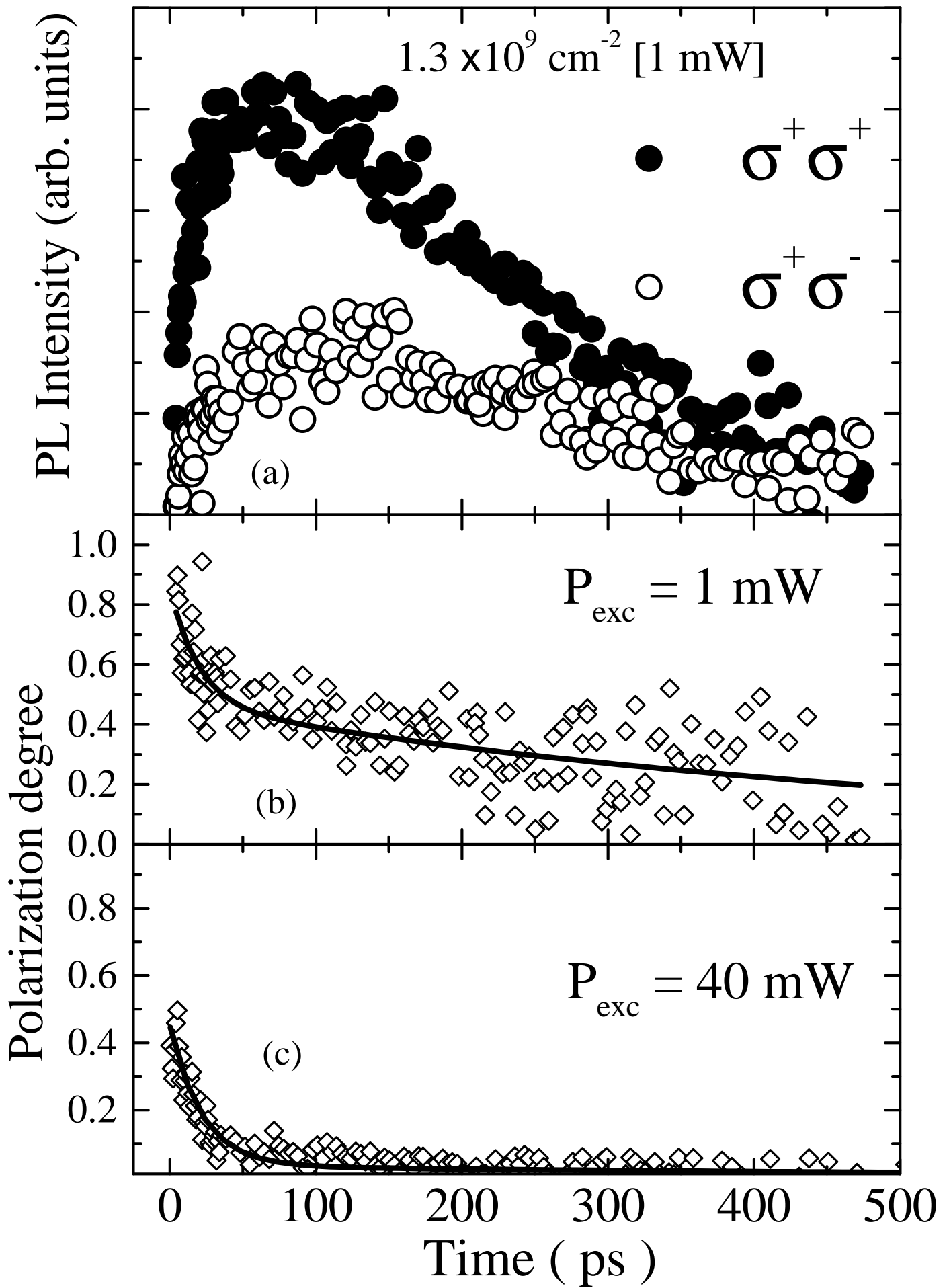

Fig. 9 (Viña)

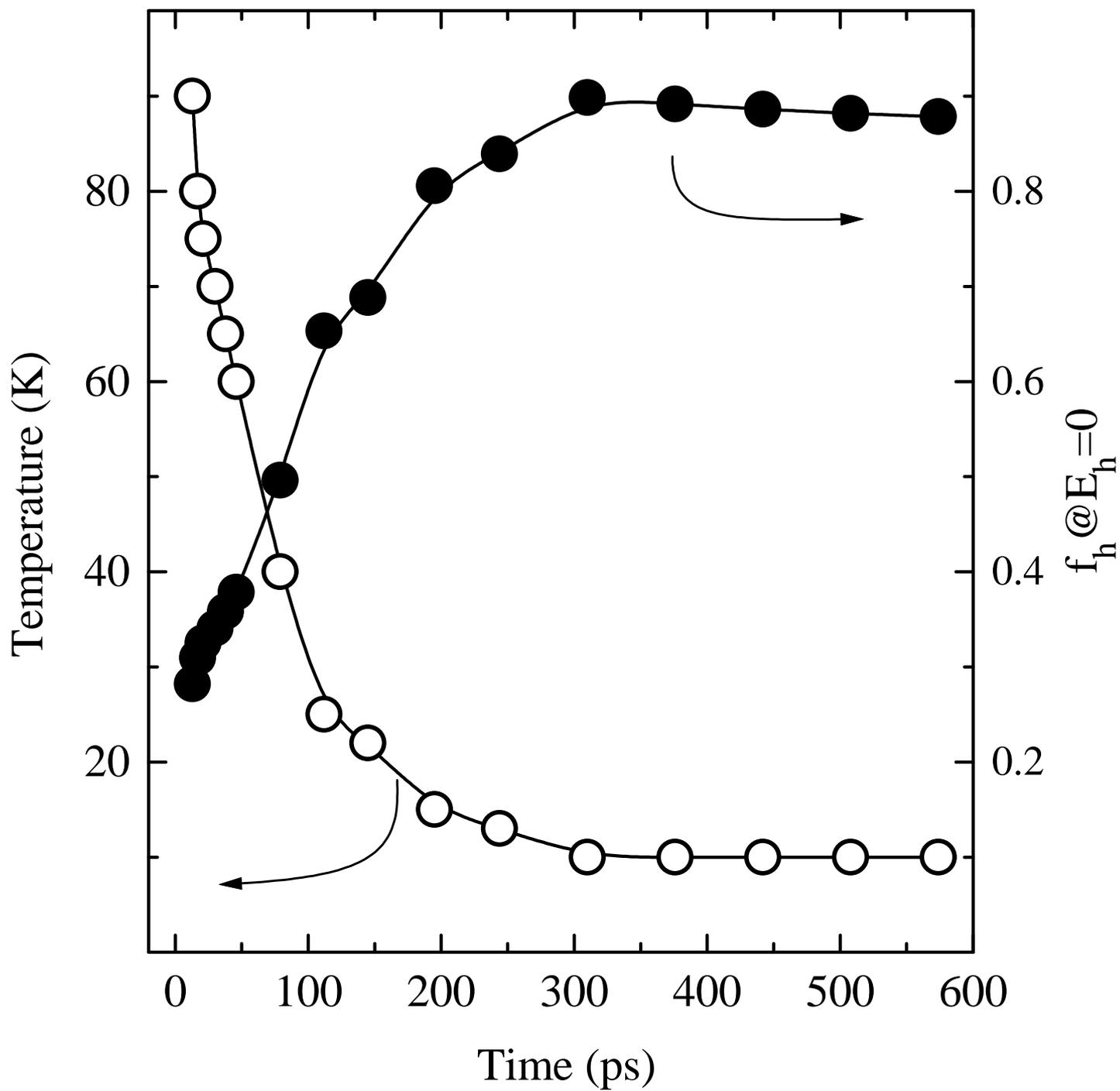

Fig. 10 (Viña)

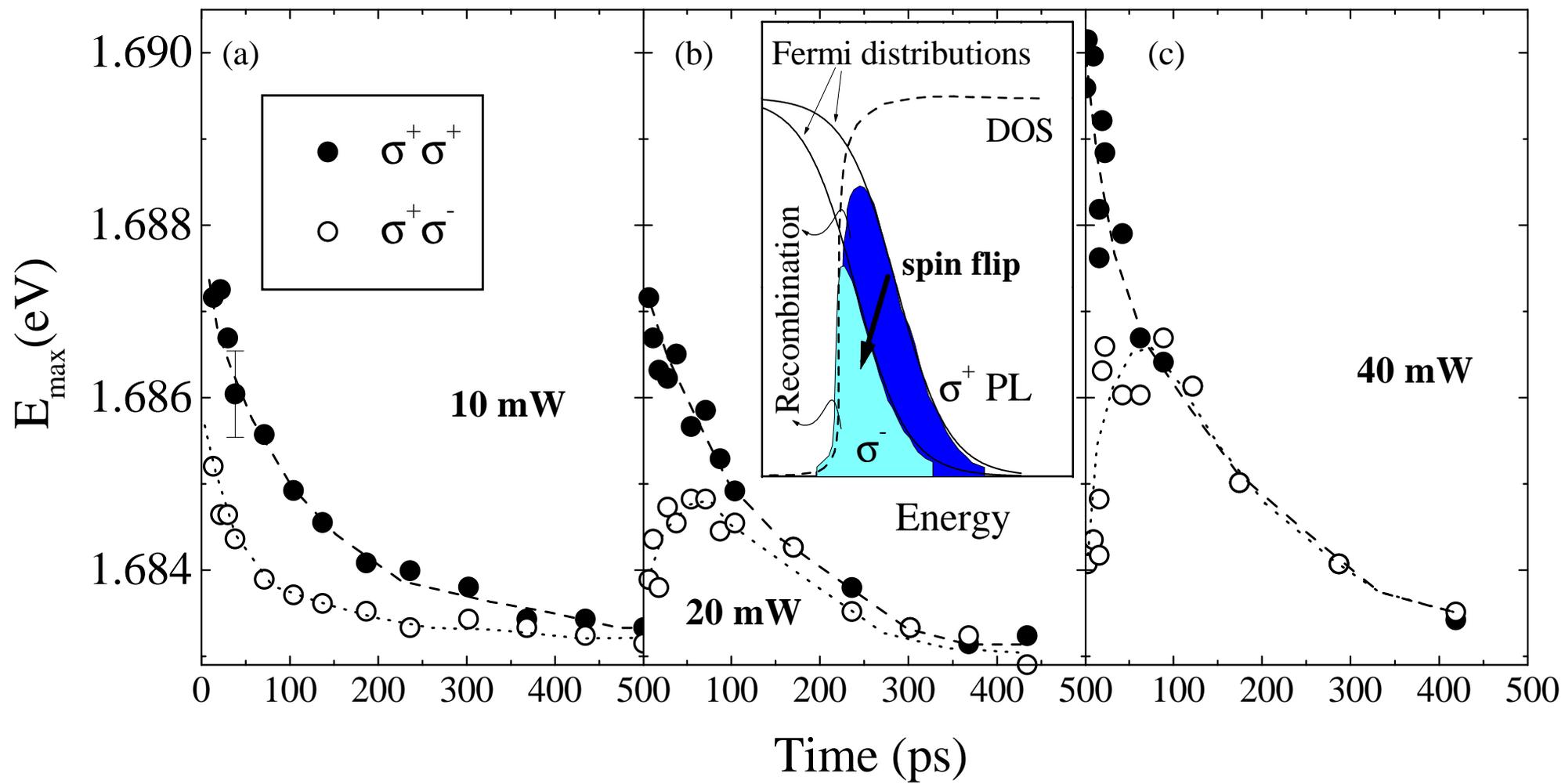

Fig. 11 (Viña)

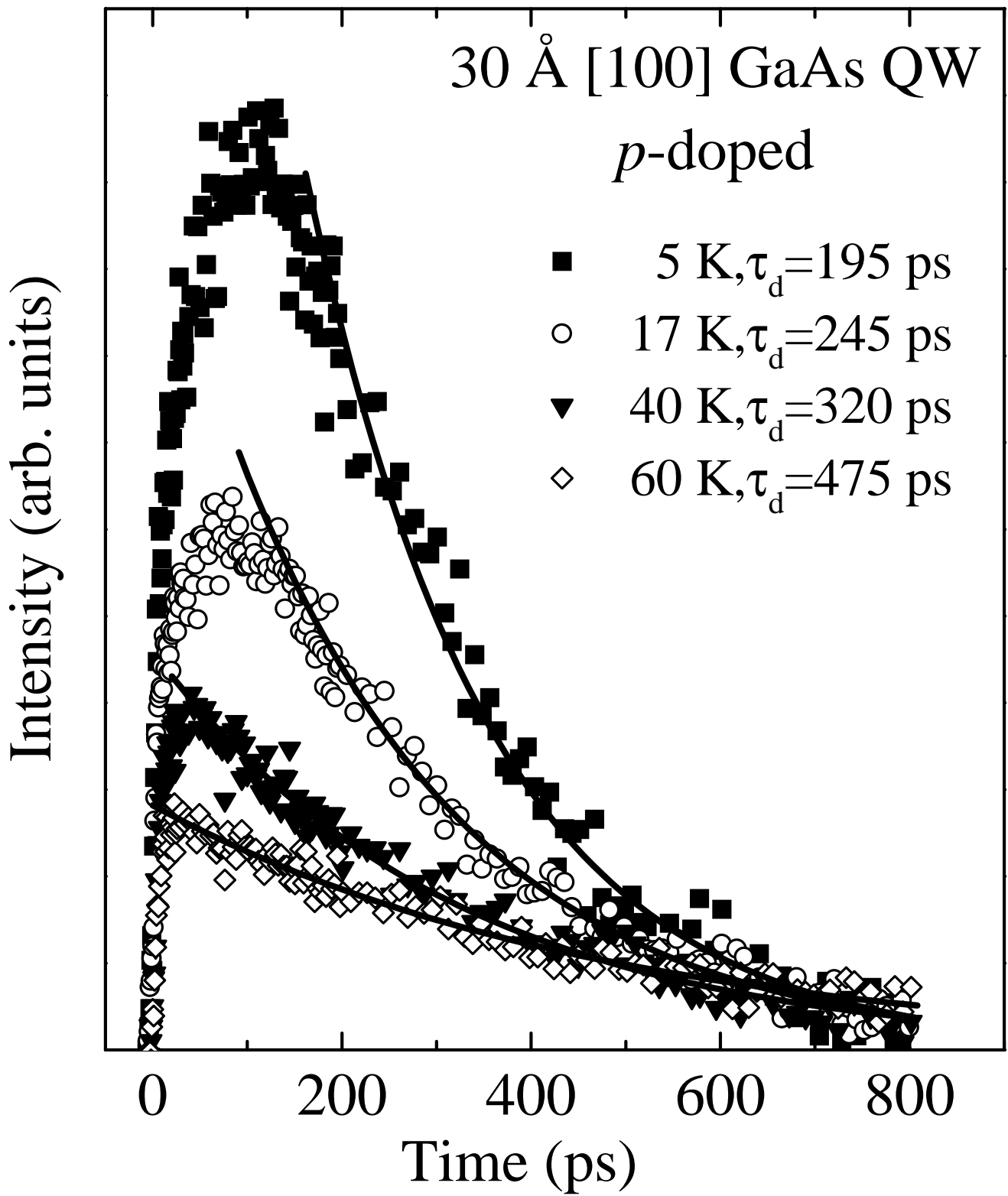

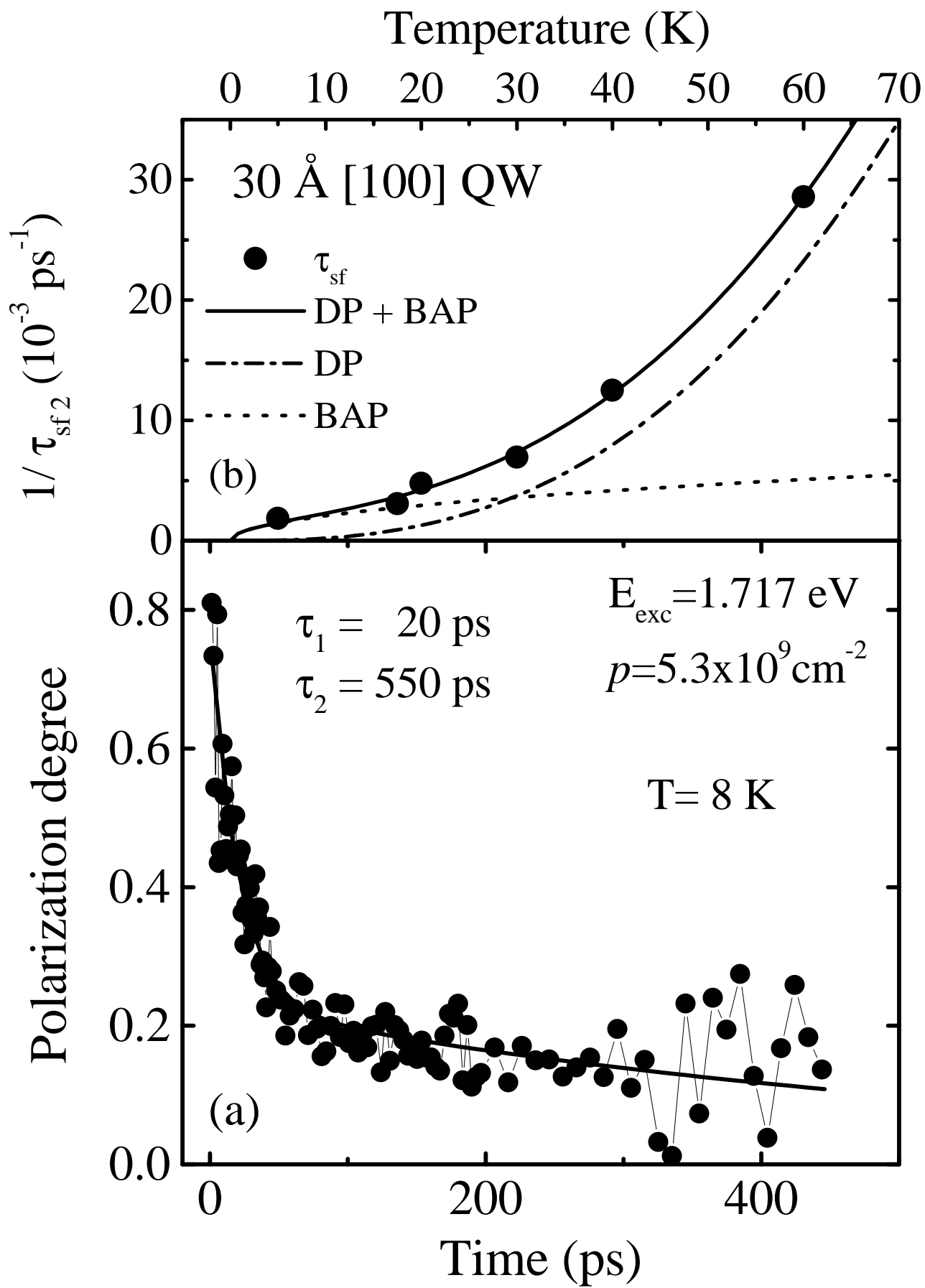

Fig.13 (Viña)